\documentclass[journal=jctcce,manuscript=article, layout=onecolumn]{achemso}

\usepackage{chemformula} 
\usepackage[T1]{fontenc} 
\setkeys{acs}{maxauthors = 18}
\usepackage{cellspace} 
\author{Luca Nils Philipp}
\email{luca-nils.philipp@uni-wuerzburg.de}
\affiliation{Institute of Physical and Theoretical Chemistry, University of Würzburg, 97074 Würzburg, Germany
}%
\author{Tomislav Begu\v{s}i\'{c}}
\email{tomislav.begusic@uni-wuerzburg.de}
\affiliation{Institute of Physical and Theoretical Chemistry, University of Würzburg, 97074 Würzburg, Germany
}

\title{High-Order Response Functions with Duschinsky Coupling and Finite Temperature: Application to Two-Dimensional Resonance Raman Spectroscopy}

\begin{document}

\begin{tocentry}

\includegraphics[]{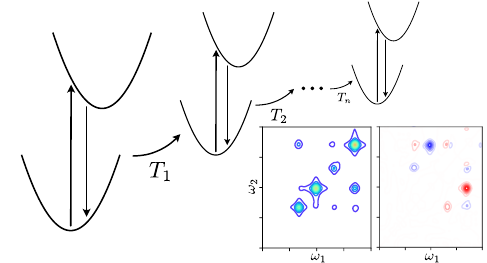}

\end{tocentry}

\begin{abstract}
In this work, we derive closed-form expressions for multi-time correlation functions underlying higher-order nonlinear spectroscopic response functions within the harmonic approximation. Our method includes displacements, frequency changes, and Duschinsky rotation between the vibrational modes of different electronic states while retaining polynomial scaling with the number of degrees of freedom. Finite temperature is accounted for without summation over vibrational eigenstates. To show the capabilities of our method, we apply it to calculate fifth-order two-dimensional resonance Raman (2DRR) spectra of two-mode model systems and naphthalene. Comparison with the displaced harmonic oscillator model shows that Duschinsky rotation modifies the positions and relative intensities of diagonal and cross peaks. 
Overall, our results demonstrate that 2DRR spectroscopy is particularly sensitive to changes in normal coordinates between electronic states and can provide distinct signatures of Duschinsky coupling in molecular systems.
\end{abstract}

\section{Introduction}

Coherent multidimensional spectroscopy offers a wealth of information about ultrafast dynamical processes in molecular systems.\cite{Cho2008Coherent,Scholes_review:2022} Two-dimensional electronic spectroscopy (2DES), for example, reveals not only electronic information, such as couplings between different electronic states, but also vibrational information that is reflected in the spectral lineshape and population dynamics.\cite{Butkus_Abramavicius:2012,book_Mukamel:1999} Electronic transitions are often employed in impulsive experiments to probe vibrational states by measuring vibrational coherence, such as in the third-order impulsive stimulated Raman scattering (ISRS)\cite{Chesnoy_Mokhtari:1988} or fifth-order time-resolved (or 2D) ISRS\cite{Kuramochi_Tahara:2019,Fumero_Scopigno:2020, Kuramochi2021, Yoneda_Kuramochi:2023} and 2D resonance Ramane (2DRR) spectroscopies.\cite{Ren_Mukamel:2013,Ren_Mukamel:2013a,Molesky_Moran:2014, Guo_Moran:2015, Molesky_Moran:2016a, Molesky_Moran:2016,Guo_Moran:2017, Cheshire_Moran:2019}

Modeling vibrational dynamics initiated by electronic transitions has a long history, with approaches ranging from numerically exact quantum dynamics on precomputed potential energy surfaces\cite{book_MCTDH} to on-the-fly quantum or mixed quantum-classical methods.\cite{Gonzlez_book:2025} In practice, the first step towards simulating and interpreting vibrationally resolved electronic spectra is the harmonic approximation.
Techniques for evaluating harmonic Franck--Condon spectra are regularly applied to absorption, emission, photoelectron,\cite{Lami_Santoro:2004,Santoro_Barone:2007,Santoro_Barone:2008,Biczysko_Barone:2009,Barone_Santoro:2009,Ferrer_Santoro:2013,Cerezo_Santoro:2013,Baiardi_Barone:2013,Tapavicza_Sundholm:2016,Souza_Izsak:2018,Souza_Izsak:2019a,Tapavicza:2019} or resonance Raman spectroscopy,\cite{Santoro_Barone:2011,Banerjee_Saalfrank:2012,AvilaFerrer_Santoro:2013,Baiardi_Barone:2014,Egidi_Barone:2014,Souza_Izsak:2019} and are implemented in standard quantum chemistry packages.\cite{g16,ORCA2020,TURBOMOLE2023,QChem5} These allow for arbitrary harmonic potential energy surfaces, accounting for differences in vibrational frequencies and normal modes (Duschinsky coupling\cite{Duschinsky:1937}) between electronic states. In contrast, in time-resolved spectroscopy, the simplest displaced harmonic oscillator (DHO) model, which neglects Duschinsky rotation of normal modes or frequency changes, is the default model. This is partly because the DHO model can be solved exactly within the second-order cumulant approach and results in a particularly simple theory that easily extends to arbitrary order response functions.\cite{book_Mukamel:1999} Nonetheless, Duschinsky effects have been identified in experiments, especially in those that probe vibrational coherence generated by ultrafast visible or UV pulses, including time-resolved pump-probe detection of resonance Raman spectra\cite{Fuji_Kobayashi:2000} and 2D-ISRS.\cite{Fumero_Scopigno:2020} In addition, the advances in broadband 2DES with vibronic resolution\cite{Bizimana_Turner:2015,Kim_Cho:2020} and hybrid electronic-vibrational (2DEV)\cite{Wu_Fleming:2019,Arsenault_Fleming:2021} and vibrational-electronic (2DVE)\cite{Courtney_Khalil:2015,Gaynor_Khalil:2017,Masood_Bredenbeck:2026} spectroscopies further motivate the need for accurate modeling of normal mode rotations.

Theoretical studies that go beyond the DHO model and treat the problem in full dimensionality are scarce. Anharmonic effects were investigated using one- or two-dimensional models, with application to 2DES\cite{Andre_Hansen:2018} and 2DRR.\cite{Guo_Moran:2015,Guo_Moran:2017} Regarding full-dimensional simulations, a group of authors studied Duschinsky and anharmonicity effects approximately by employing third-order cumulant expansion in the context of condensed-phase simulations, i.e., in an explicit solvent environment.\cite{Zuehlsdorff_Isborn:2020,Zuehlsdorff_Isborn:2021} Concurrently, a Gaussian wavepacket method, the on-the-fly ab initio thawed Gaussian approximation,\cite{Wehrle_Vanicek:2014, Wehrle_Vanicek:2015,Begusic_Vanicek:2020a} was employed to improve upon the DHO model. Since the Gaussian wavepacket dynamics is exact for arbitrary harmonic potential energy surfaces, this method simulates Duschinsky effects in 2DES exactly at a polynomial cost $\mathcal{O}(N^3)$ in the number of degrees of freedom $N$. This was then extended to finite temperature\cite{Begusic_Vanicek:2021} using the concept of thermofield dynamics,\cite{Suzuki:1985,Begusic_Vanicek:2020} leaving the polynomial scaling unchanged and avoiding the sum-over-states formula for the Boltzmann density operator. Within the harmonic approximation, both time-dependent parameters of the Gaussian wavepacket and wavepacket overlaps have closed-form solutions. Therefore, a full set of closed-form equations is available, rendering numerical propagation schemes or approximate treatments of Duschinsky effects unnecessary. This has been derived using the quantum optics language of squeezed coherent states and rotation, displacement, and squeeze operators.\cite{Quintela_Troiani:2023} Alternatively, closed-form solutions for third-order response functions have been derived using the path-integral formalism.\cite{Allan_Zuehlsdorff:2025} The latter has been traditionally used to derive closed-form expressions of time-correlation functions for steady-state spectroscopies and is the one we employ in this work. The same authors expanded their formulation, named Franck--Condon 2DES (FC2DES), to Herzberg-Teller effects (FC2DES+HT),\cite{Allan_Zuehlsdorff:2025a} thus elevating the computation of third-order response functions to the same level of theory as standard steady-state spectroscopy. However, that work stopped at the third order, with no closed-form expressions or codes available for the simulation of higher-order response functions.

Here, we derive closed-form equations for arbitrary-order, multi-time correlation functions and show that these can be obtained elegantly in a way analogous to the derivations of one-time and three-time correlation functions. Unsurprisingly, high-order correlation functions still admit the same polynomial-time scaling in the number of degrees of freedom, as well as an additional polynomial (asymptotically linear) scaling with the order $n$, which is in line with the cost of simulating squeezed coherent state dynamics.\cite{Quintela_Troiani:2023} We then apply the method to study Duschinsky effects in the fifth-order 2DRR spectra of model systems and naphthalene.

\section{Methods}

\subsection{High-Order Response Functions}

In any kind of nonlinear spectroscopic technique, a sample interacts multiple times with an external electric field $E(t)$ composed of several light pulses. Within the sample, $n$ interactions with this external field lead to the creation of a time-dependent $n$th-order nonlinear polarization
\begin{equation}
        P^{(n)}(t) = \int_0^\infty dt_n\int_0^\infty dt_{n-1} \cdots \int_0^\infty dt_{1} 
        E(t-t_n)E(t-t_n-t_{n-1})\cdots E(t-t_n-\cdots-t_1) R^{(n)}(t_1, \dots, t_n)
\end{equation}
where $t_1,t_2,\dots, t_n$ are the times at which interactions occur and $R^{(n)}(t_1, \cdots, t_n)$ is the \textit{n}th order response function. Subsequently, the polarization is emitted and can be detected with various methods. The response of the sample is encoded in the \textit{n}th order response function given by
\begin{eqnarray}
        R^{(n)}(t_1, \dots, t_n) &=& \left(\frac{i}{\hbar}\right)^n \theta(t_1) \theta(t_2) \dots \theta(t_n) \nonumber \\
        &\times& \mathrm{Tr}\{\mu(t_1+\cdots+t_n) 
    \left[\mu(t_1+\cdots+t_{n-1}), \left[\dots\left[\mu(t_1),\left[\mu(0),\rho_\mathrm{eq}\right]\right]\dots\right]\right]\},
\end{eqnarray}
with the Heaviside function $\theta(t)$, the dipole operator within the Heisenberg picture $\mu(t)$, and the density matrix in thermal equilibrium $\rho_\mathrm{eq}$. Due to the nested commutators in the previous equation, various terms contribute to the response function, all of which can be cast into the following general form
\begin{equation}
    R_i^{(n)}(t_1,\dots, t_n)=\mathrm{Tr}[\mu(t'_n)\cdots\mu(t'_1)\mu(t'_0)\rho_\mathrm{eq}].
\end{equation}
For each $i$, $(t'_{0},t'_1,\dots,t'_n)$ is a permutation of $(0,t_1, t_1+t_2,\dots, t_1+t_2+\cdots+t_n)$. The different $R_i^{(n)}(t_1,\dots,t_n)$ terms are commonly referred to as Liouville space pathways.

To simplify the notation, at this stage we consider a system consisting of two electronic states, the ground state $|g\rangle$ and a single electronically excited state $|e\rangle$. This will be generalized later. The vibrational degrees of freedom in the ground and excited states are described by the Hamiltonians $H_g$ and $H_e$, respectively. We further assume the Condon approximation for the transition dipole moment between the ground and the excited electronic states, $\mu_{\text{eg}}(q) \approx \text{const.}$

Under these assumptions, the general form of the Liouville space pathways can be written as
\begin{eqnarray}\label{pathways}
    R_i^{(n)}(t_1,\dots, t_n) &=& |\mu_{eg}|^{n+1} \exp\left[i \omega_{eg} \sum_{j=0}^n(-1)^j t_j'\right] \nonumber\\
    &&\times C(t'_0-t'_n-i\beta, t'_1-t'_0, t'_2-t'_1, \dots, t'_n-t'_{n-1}) \\
    C(\tau_0, \tau_1, \dots, \tau_n) &=& \frac{1}{Z}\mathrm{Tr}(e^{-iH_e\tau_n}e^{-iH_g\tau_{n-1}}\cdots e^{-iH_e\tau_{1}}e^{-iH_g\tau_{0}}), \label{pathways_C}
\end{eqnarray}
where $\omega_{eg}$ is the adiabatic excitation frequency, corresponding to the energy gap between the minima of the ground- and excited-state potential energy surfaces, and $Z$ is the partition function corresponding to the ground-state vibrational Hamiltonian, $Z=\mathrm{Tr}\left(e^{-\beta H_g}\right)$. In the following, we derive analytic expressions for $C(\tau_0, \tau_1, \dots, \tau_n)$ for the case when the ground- and excited-state potential energy surfaces are approximated by general multimode harmonic potentials.

The ground- and excited-state vibrational Hamiltonians are given by
\begin{equation}
    H_g = \frac{1}{2}p_g^Tp_g + \frac{1}{2}q_g^T\Omega^2_gq_g
\end{equation}
and 
\begin{equation}
    H_e = \frac{1}{2}p_e^Tp_e + \frac{1}{2}q_e^T\Omega^2_eq_e,
\end{equation}
where $q_g$ ($q_e$) are the coordinates of the ground-state (excited-state) normal modes, $p_g$ ($p_e$) are the respective momenta, and $\Omega_g$ ($\Omega_e$) is a matrix with the frequencies of ground-state (excited-state) normal modes on the diagonal. Ground- and excited-state normal-mode coordinates are assumed to be related to each other by a linear transformation
\begin{equation}\label{duschinsky}
    q_g = Jq_e + K
\end{equation}
with the displacement vector $K$ and the Duschinsky rotation matrix $J$.

To obtain a closed-form expression for a general Liouville space pathway, we exploit the fact that analytical formulas for the position-representation matrix elements of the time-evolution operator $e^{-iHt}$ are well known if $H$ describes a system of harmonic oscillators. To this end, we carry out the trace in Eq.~(\ref{pathways_C}) within the continuous basis of position states of the ground-state normal coordinates and introduce a resolution of identity in the ground-state normal modes at various times
\begin{equation}\label{response start}
    \begin{aligned}
        C(\tau_0, \tau_1, \dots, \tau_n) = &\frac{1}{Z} \int dq_{g,0}\cdots \int dq_{g,n} \\ 
        &\times \langle q_{g,0}|e^{-iH_e\tau_n}|q_{g,n}\rangle\langle q_{g,n}| e^{-iH_g\tau_{n-1}}| q_{g,n-1}\rangle \cdots \\
        &\times \langle q_{g,2}|e^{-iH_e\tau_{1}}|q_{g,1}\rangle\langle q_{g,1}|e^{-iH_g\tau_{0}}|q_{g,0}\rangle.
    \end{aligned}
\end{equation}
To resolve the matrix elements of the excited-state time-evolution operator in the ground-state coordinates, we rewrite it as
\begin{equation}\label{excited propagator in ground coordinates}
    \langle q_g|e^{-iH_et}|q_g'\rangle = \int dq_e\int dq_e' \langle q_g|q_e\rangle\langle q_e|e^{-iH_et}|q_e'\rangle\langle q_e'|q_g'\rangle.
\end{equation}
Since the normal modes of ground and excited states are related to each other by a linear transformation (\ref{duschinsky}), the normal-mode position states satisfy
\begin{equation}\label{duschinsky basis}
    \langle q_{e}|q_{g}\rangle = \delta(q_g - (Jq_e+K)).
\end{equation}
Furthermore, the matrix element of the quantum harmonic oscillator time-evolution operator is
\begin{equation}\label{propagator}
    \langle q_\alpha|e^{-iH_\alpha\tau}|q'_\alpha\rangle = \sqrt{\frac{\det(a_\alpha(\tau))}{(2\pi i)^{N}}} G(q_\alpha, q'_\alpha, a_{\alpha}(\tau), b_{\alpha}(\tau))
\end{equation}
where
\begin{equation}
    G(q, q', a, b) = \exp\left(\frac{i}{2}q^T b q + \frac{i}{2}q'^T b q' - iq^T a q'\right),
\end{equation}
is a Gaussian in coordinates $q$ and $q'$, $\alpha\in\{g,e\}$, and $a_\alpha(\tau)$, $b_\alpha(\tau)$ are diagonal $N$-dimensional matrices
\begin{eqnarray}
    a_\alpha(\tau)&=&\Omega_\alpha \sin(\Omega_\alpha \tau)^{-1}, \\
    b_\alpha(\tau)&=&\Omega_\alpha \tan(\Omega_\alpha \tau)^{-1}.
\end{eqnarray}
Inserting Eqs. (\ref{excited propagator in ground coordinates})-(\ref{propagator}) into Eq.~(\ref{response start}) followed by integration over all ground-state normal coordinates yields

\begin{equation}\label{response intermediate}
    \begin{aligned}
        C(\tau_0, \tau_1, \dots, \tau_n) = &\frac{1}{Z} \sqrt{\frac{\det(a_g(\tau_0)a_e(\tau_1)\cdots a_g(\tau_{n-1})a_e(\tau_{n}))}{(2\pi i)^{N(n+1)}}} \int dq_{e,0}\cdots \int dq_{e,n} \\
        &\times G(q_{e, n}, q_{e, n-1}, a_e(\tau_n), b_e(\tau_n)) \\
        &\times G(Jq_{e, n-1}+K, Jq_{e, n-2}+K, a_g(\tau_{n-1}), b_g(\tau_{n-1}))\dots \\
        &\times G(q_{e, 1}, q_{e, 0}, a_e(\tau_1), b_e(\tau_1))\\
        &\times G(Jq_{e, 0}+K, Jq_{e, n}+K, a_g(\tau_0), b_g(\tau_0)),
    \end{aligned}
\end{equation}
We now note that the integrand is a product of Gaussian functions, i.e., it is a Gaussian in the joint, $(n+1)N$-dimensional coordinate
\begin{equation}
    z^T = \Big(q_{e,0}^T \quad q_{e,1}^T \quad \cdots \quad q_{e,n}^T\Big).
\end{equation}
More precisely,
\begin{equation}\label{response almost final}
\begin{aligned}
    C(\tau_0, \tau_1, \dots, \tau_n) = &\frac{1}{Z} \sqrt{\frac{\det(a_g(\tau_0)a_e(\tau_1)\cdots a_g(\tau_{n-1})a_e(\tau_{n}))}{(2\pi i)^{N(n+1)}}} \\
    &\times \exp\left[iK^T \tilde{c} K\right] \int dz \exp\left(\frac{i}{2}z^TDz + iE^Tz\right),
\end{aligned}
\end{equation}
where
$D$ is a $(n+1)N \times (n+1)N$ block-tridiagonal matrix
\begin{equation}
\label{eq:D}
    D = \begin{pmatrix}
        B_0 & A_1 &  &  & A_0\\
        A_1 & B_1 & A_2 &  & \\
         & A_2 & \ddots & \ddots & \\
         &  & \ddots & B_{n-1} & A_n\\
        A_0 &  &  & A_n& B_n 
    \end{pmatrix}
\end{equation}
with
\begin{eqnarray}
    A_i = \begin{cases}
        -\bar{a}_g(\tau_i),& i \text{ even} \\
        -a_e(\tau_i), & i \text{ odd}
    \end{cases}, \\
    B_i =\begin{cases}
        \bar{b}_g(\tau_i) +b_e(\tau_{i+1}), & i \text{ even} \\
        b_e(\tau_i) + \bar{b}_g(\tau_{i+1}), & i \text{ odd}
    \end{cases}
\end{eqnarray}
and
\begin{eqnarray}
    \bar{a}_g(\tau_i) &=& J^Ta_g(\tau_i)J, \\
    \bar{b}_g(\tau_i) &=& J^Tb_g(\tau_i)J, \\
    c(\tau_i) &=& b_g(\tau_i) - a_g(\tau_i), \\
    \tilde{c} &=& \left(\sum_{k=0}^{(n-1)/2}c(\tau_{2k})\right)
\end{eqnarray}
$E$ is a $(n+1)N$-dimensional vector
\begin{eqnarray}
    E^T &=& \left(E_0^T\quad E_1^T\quad \dots\quad E_n^T \right), \label{eq:E} \\
    E_i &=& J^T c(\tau_{\lfloor i / 2 \rfloor}) K.
\end{eqnarray}
The analytical formula for the multidimensional Gaussian integral is
\begin{equation}
    \int dz \exp\left(\frac{i}{2}z^TDz + iE^Tz\right) = \sqrt{\frac{(2\pi i)^{N(n+1)}}{\det(D)}}\exp\left(-\frac{i}{2}E^TD^{-1}E\right).
\end{equation}
Inserting this equation into Eq.~(\ref{response almost final}) leads to the final expression for $n$th order correlation function:
\begin{equation}\label{response final}
\begin{aligned}
    C(\tau_0, \tau_1, \dots, \tau_n) = &\frac{1}{Z} \sqrt{\frac{\det(a_g(\tau_0)a_e(\tau_1)\cdots a_g(\tau_{n-1})a_e(\tau_{n}))}{\det(D)}} \\
    &\times \exp\left(iK^T\tilde{c}K-\frac{i}{2}E^TD^{-1}E\right).
\end{aligned}
\end{equation}

The prefactor in Eq.~(\ref{response final}) is evaluated in the logarithmic representation, which ensures that we do not divide very large or very small numbers. Furthermore, the inverse of $D$ is never explicitly evaluated. Instead, only $D^{-1}E$ is computed by a linear solver. To obtain continuous response functions, we unwrap the phase of the calculated response functions. In general, the computational cost of our implementation is $\mathcal{O}(n^3N^3)$. We note, however, that the asymptotic scaling of $\mathcal{O}(nN^3)$ is possible by employing algorithms specifically designed for block-tridiagonal matrices. These algorithms come with additional prefactors because they involve multiple explicit matrix inverse computations of $N \times N$ matrices. Therefore, their overall computational cost is likely to be higher in the regime of large $N$ and comparably small $n$, e.g., for third- or fifth-order spectroscopy of molecular systems with $N \approx 50$. In our implementation and for applications presented here, we work with a dense $D$, keeping in mind that for other applications one could profit from this improved asymptotic scaling. Additional practical speed-up can be achieved by employing graphical processing units, as demonstrated for third-order response functions.\cite{Allan_Zuehlsdorff:2025,Allan_Zuehlsdorff:2025a} Finally, we note that this cost analysis applies to a single evaluation for a specific choice of $(\tau_0, \tau_1, \dots, \tau_n)$, i.e., for a fixed set of delay times $t_1, t_2, \dots, t_n$. For a fixed number of steps along each time axis, the total cost to map out $R_i^{(n)}(t_1, t_2, \dots, t_n)$ is exponential in $n$. In our simulation of 2DRR, we leveraged the fact that certain time delays are limited by the ultrashort pulse duration, meaning that only two experimentally detected delay times require a large number of time steps.

\subsection{General expressions for multi-state correlation functions}

For completeness, we provide explicit expressions for computing correlation functions involving more than two electronic states. These can be derived by following the same procedure as above. Let
\begin{equation}
    C(\tau_0, \tau_1, \dots, \tau_n) = \frac{1}{Z_0}\mathrm{Tr}(e^{-iH_n\tau_n}e^{-iH_{n-1}\tau_{n-1}}\cdots e^{-iH_1\tau_{1}}e^{-iH_0\tau_{0}}), \label{general_C}
\end{equation}
where the time evolution during time $\tau_i$ is governed by the harmonic vibrational Hamiltonian
\begin{equation}
    H_i = \frac{1}{2} p_i^T p_i + \frac{1}{2} q_i^T \Omega_i^2 q_i.
\end{equation}
We assume that the initial vibrational state is a thermal equilibrium state of $H_0$ with partition function $Z_0 = \text{Tr}\left(e^{-\beta H_0}\right)$ and we make no further assumption about its relation to the molecular system, i.e., the initial electronic state can be the ground or one of the excited states. Normal mode coordinates of all electronic states are assumed to be related to a reference coordinate system by a linear, Duschinsky transformation,
\begin{equation}
    q_i = J_i q_{\text{ref}} +  K_i,
\end{equation}
as done before in the two-state derivation where we had $q_\text{ref} \equiv q_e$. Given $\Omega_i$, $J_i$, and $K_i$ for $i=0,1,\dots,n$, we can evaluate the $n$-th order correlation function as
\begin{equation}\label{response general}
\begin{aligned}
    C(\tau_0, \tau_1, \dots, \tau_n) = &\frac{1}{Z_0} \sqrt{\frac{\det\left[\prod_{i=0}^na_i(\tau_i)\right]}{\det(D)}} \exp\left(i\sum_{i=0}^n K_i^T c_i K_i-\frac{i}{2}E^TD^{-1}E\right).
\end{aligned}
\end{equation}
Here, $D$ is defined according to Eq.~(\ref{eq:D}) with modified blocks
\begin{eqnarray}
    A_i &=& -J_i^T a_i(\tau_i) J_i,\\
    B_i &=& J_i^T b_i(\tau_i)J_i + J_{i+1}^T b_{i+1}(\tau_{i+1})J_{i+1}, \mathrm{with}\, n+1 \equiv 0,
\end{eqnarray}
$E$ is constructed according to Eq.~(\ref{eq:E}) with
\begin{equation}
    E_i = J_i^T c_i(\tau_i) K_i + J_{i+1}^T c_{i+1}(\tau_{i+1}) K_{i+1},
\end{equation}
and
\begin{eqnarray}
    a_i(\tau) &=& \Omega_{i} \sin(\Omega_{i} \tau)^{-1},\\
    b_i(\tau) &=& \Omega_{i} \tan(\Omega_{i} \tau)^{-1},\\
    c_i(\tau) &=& b_i(\tau) - a_i(\tau).
\end{eqnarray}

\subsection{Linear Absorption Spectroscopy}

The process of linear absorption is contained as a special case in the nonlinear theory. For linear spectroscopy, the first-order polarization $P^{(1)}(t)$ is the quantity of interest, which is given by
\begin{equation}
    P^{(1)}(t) = \int_0^\infty dt_1 E(t-t_1)R^{(1)}(t_1)
\end{equation}
with the first-order response function
\begin{equation}
    R^{(1)}(t) = \frac{i}{\hbar} \theta(t) \text{Tr}\{\hat{\mu}(t)[\hat{\mu}(0),\rho_{\rm eq}]\} = \frac{i}{\hbar} \theta(t) [R_1^{(1)}(t) - R_1^{(1)}(t)^*],
\end{equation}
In the second equality, we assume the Condon approximation and label the Liouville space pathway
\begin{equation}
    R_1^{(1)}(t) = |\mu_{eg}|^2e^{-i\omega_{eg}t}C(-t-i\beta, t),
\end{equation}
where $C(\tau_0, \tau_1)$ is defined according to Eq.~(\ref{pathways_C}). Then the absorption cross-section
\begin{equation}\label{sigma_omega}
    \sigma(\omega) = \frac{4\pi\omega}{c} \frac{\text{Im}[\tilde{P}^{(1)}(\omega) \tilde{E}(\omega)^*]}{|\tilde{E}(\omega)|^2}
\end{equation}
evaluates to
\begin{equation}\label{linear_absorption}
    \sigma(\omega) = \frac{4\pi\omega}{\hbar c} \text{Re}\left[ \int_0^{\infty} dt e^{i\omega t} R_1^{(1)}(t)\right],
\end{equation}
where in Eq.~(\ref{sigma_omega}) we used the Fourier transform convention $\tilde{f}(\omega) = \int_{-\infty}^{\infty}dt\exp(i\omega t) f(t)$ and in Eq.~(\ref{linear_absorption}) we dropped the counter-rotating term that is zero for a system initially in the ground electronic state.




\subsection{2D Resonance Raman Spectroscopy}

\begin{figure}[pth]
  \includegraphics[]{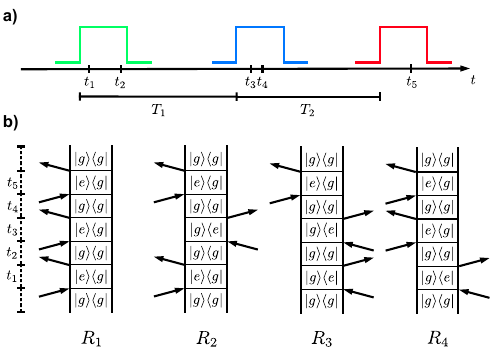}
  \caption{\label{figure1}a) Pulse sequence producing the 5\textit{th} order polarization, which is used to extract the 2DRR spectra. b) Liouville space pathways connected to the ground state 2DRR signal.}
\end{figure}

Experimentally, 2D resonance Raman spectra are recorded using a sequence of three laser pulses, separated by delay times $T_1$ and $T_2$. The corresponding electric field can be expressed as
\begin{equation}
    E(t) = E_1(t+T_1 + T_2) + E_2(t+T_2) + E_3(t),
\end{equation}
where we set the final pulse at time zero. For simplicity, we choose to model each field as a rectangular pulse (see Fig.~\ref{figure1}a) of duration $\Delta t$,
\begin{eqnarray}
    E_i(t) &=& \mathcal{E}(t) + \mathcal{E}(t)^{\ast}, \\
    \mathcal{E}(t) &=& \mathcal{E}_0 e^{-i\omega_L t} \text{rect} \left( \frac{t}{\Delta t}\right),
\end{eqnarray}
where $\mathcal{E}_0$ is the amplitude, $\omega_L$ is the carrier frequency, and
\begin{equation}
    \text{rect}\left(\frac{t}{\Delta t}\right) = \begin{cases}
        1, \quad 0\le t \le \Delta t, \\
        0, \quad \text{otherwise}
    \end{cases}.
\end{equation}
The system interacts twice with each pulse to create vibrational coherences in the ground or excited electronic state during the two delay times $T_1$ and $T_2$. Assuming non-overlapping pulses, phase-matching conditions, and heterodyne detection using the final light pulse, we can express the signal as
\begin{equation}
    S^{(5)}(T_1, T_2) = \text{Im} \int_{-\infty}^{\infty} dt \mathcal{E}(t)^{\ast} P^{(5)}(t; T_1, T_2),
\end{equation}
where
\begin{equation}
\begin{aligned}
    P^{(5)}(t; T_1, T_2) = &\int_0^\infty dt_1 \int_0^\infty dt_2 \int_0^\infty dt_3 \int_0^\infty dt_4 \int_0^\infty dt_5 \mathcal{E}(t-t_5) E_2(t-t_5-t_4+T_2)\\ 
    &\times E_2(t-t_5-t_4-t_3+T_2) E_1(t-t_5-t_4-t_3-t_2+T_1+T_2)\\
    &\times E_1(t-t_5-t_4-t_3-t_2-t_1+T_1+T_2) R^{(5)}(t_1,t_2,t_3,t_4,t_5)
    \label{eq:polarization5th}
\end{aligned}
\end{equation}
is the time-dependent polarization of the system generated through nonlinear interaction with the light pulses. Here, we focus on pathways in $R^{(5)}$ that remain in the ground electronic state during times $t_2$ and $t_4$ (see Fig.~\ref{figure1}b, which are analogous to the off-resonance 2D Raman spectroscopy.\cite{Tanimura_Mukamel:1993,Palese_Miller:1994} Experimentally, these pathways could be selected using the preresonance regime in which the frequency of the laser pulses is close to but below the frequency of the electronic transition.\cite{book_Mukamel:1999} Then, only 4 out of 16 pathways remain, together with their complex conjugates:
\begin{equation}
\begin{aligned}
    R^{(5)}(t_1,t_2,t_3,t_4,t_5) = &\left(\frac{i}{\hbar}\right)^5 \theta(t_1)\theta(t_2)\theta(t_3)\theta(t_4)\theta(t_5)\\
    & \times \sum_{i=1}^4\Big[R_i^{(5)}(t_1,t_2,t_3,t_4,t_5) - R_i^{(5)}(t_1,t_2,t_3,t_4,t_5)^*\Big]. \label{eq:R5th}
\end{aligned}
\end{equation}
Within the rotating-wave approximation, the complex conjugate terms, which come with a phase of $e^{i\omega_{eg} t_5}$, are neglected and each pathway is accompanied by a choice of $\mathcal{E}$ or $\mathcal{E}^\ast$ for $E_1$ and $E_2$ appearing in Eq.~(\ref{eq:polarization5th}). For example, the contribution of $R^{(5)}_1$ to the final signal is
\begin{equation}
\begin{aligned}
    &S_1^{(5)}(T_1, T_2) = \text{Im}\int_{-\infty}^{\infty} dt \int_0^{\infty} dt_5 \dots \int_0^{\infty} dt_1 \mathcal{E}(t)^\ast  \mathcal{E}(t-t_5) \\
    &\times \mathcal{E}(t-t_5-t_4+T_2)^\ast \mathcal{E}(t-t_5-t_4-t_3+T_2) \\
    &\times \mathcal{E}(t-t_5-t_4-t_3-t_2+T_1+T_2)^\ast \mathcal{E}(t-t_5-t_4-t_3-t_2-t_1+T_1+T_2)\\
    &\times \left(\frac{i}{\hbar}\right)^5 R_1^{(5)}(t_1,t_2,t_3,t_4,t_5).
    \label{eq:S_1_full}
\end{aligned}
\end{equation}
We further simplify this expression by assuming pulses that are short compared to the typical vibrational period and compared to delay times $T_1$ and $T_2$. Then, we can identify $t_2 \approx T_1$ and $t_4 \approx T_2$, and Eq.~(\ref{eq:S_1_full}) reduces to
\begin{equation}
    S_1^{(5)}(T_1, T_2) = \frac{\mathcal{E}_0^6}{\hbar^5} \text{Re} \int_0^{\Delta t} dt_5 \int_0^{\Delta t} dt_3 \int_0^{\Delta t} dt_1 e^{i\omega_L(t_1 + t_3 + t_5)}R_1^{(5)}(t_1,T_1,t_3,T_2,t_5).
\end{equation}
The full signal is
\begin{equation}
    \label{eq:S5_final}
    S^{(5)}(T_1, T_2) = \frac{\mathcal{E}_0^6}{\hbar^5} \text{Re} \int_0^{\Delta t} dt_5 \int_0^{\Delta t} dt_3 \int_0^{\Delta t} dt_1 \sum_{j=1}^4 e^{i\omega_L[\vartheta_j(t_1, t_3) + t_5]}R_j^{(5)}(t_1,T_1,t_3,T_2,t_5),
\end{equation}
$R^{(5)}_1$--$R^{(5)}_4$ are evaluated according to Eqs.~\ref{pathways}--\ref{pathways_C} and $\vartheta_j$ are the appropriate sums or differences of $t_1$ and $t_3$, as listed in Table~\ref{table1}.

\begin{table}[]
\caption{Definition of the time arguments in Eq.~(\ref{pathways_C}) of the four different Liouville space pathways related to the ground-state 2DRR signal and $\vartheta_j(t_1, t_3)$ of Eq.~(\ref{eq:S5_final}).}
\begin{tabular}{c|ccccccc}\label{table1}
 & $\tau_5$ & $\tau_4$ & $\tau_3$ &  $\tau_2$ & $\tau_1$  &  $\tau_0$ & $\vartheta$ \\ \hline
R1 & $t_5$ & $t_4$ & $t_3$ & $t_2$ & $t_1$ & $-t_1-t_2-t_3-t_4-t_5-i\beta$ & $t_1 + t_3$ \\
R2 & $-t_3$ & $-t_4-t_5$ & $t_5$ & $t_2+t_3+t_4$ & $t_1$ & $-t_1-t_2-i\beta$ & $t_1 - t_3$ \\
R3 & $-t_1$ & $-t_2$ & $-t_3$ & $-t_4-t_5$ & $t_5$ & $t_1+t_2+t_3+t_4-i\beta$ & $-t_1 - t_3$\\
R4 & $-t_1$ & $-t_2-t_3-t_4-t_5$ & $t_5$ & $t_4$ & $t_3$ & $t_1+t_2-i\beta$ & $-t_1 + t_3$
\end{tabular}
\end{table}

Finally, the 2DRR spectrum is obtained by carrying out a Fourier transformation with respect to the two delay times and taking the absolute value,
\begin{equation}
    \sigma_\text{2DRR}(\omega_1,\omega_2) = \left|\int_0^{\infty}\int_0^{\infty} S^{(5)}(T_1, T_2)e^{-i\omega_1 T_1}e^{-i\omega_2 T_2}dT_1 dT_2\right|.
\end{equation}

\subsection{Computational Details}\label{computational}
We test our methodology by calculating linear absorption and 2DRR spectra of two test systems: two-mode model Hamiltonians and the naphthalene molecule. The two-mode model Hamiltonians are used to investigate the influence of Duschinsky coupling on linear absorption and 2DRR spectra. Naphthalene, as a medium-sized organic molecule, represents a test system in which the influence of Duschinsky coupling on the linear absorption spectra has been well studied. There, we analyze how the signatures of Duschinsky coupling in the two-mode model Hamiltonians transfer to real molecules.

The parameters of the two-mode model Hamiltonians can be found in Table~\ref{table2}. Optimizations and frequency calculations of the ground and first excited state of naphthalene were carried out in the framework of (TD-)DFT utilizing the PBE0 functional with the def2-TZVP basis set implemented in the ORCA 6 software package.\cite{https://doi.org/10.1002/wcms.70019, 10.1063/1.472933, 10.1039/b508541a} The Tamm-Dancoff approximation was applied for all excited-state calculations. From these calculations we obtained the shift vector ($K$) and the Duschinsky matrix ($J$) as outlined by Baiardi \textit{et al.}\cite{Baiardi_Barone:2013}. 

When calculating spectra using response functions, it is important to choose the size and spacing of the time grids carefully to assure that the Fourier transformations are well-behaved. For the calculation of all linear spectra we chose a grid containing 400 points reaching up to 1000~fs. Broadening of the linear spectra was introduced by multiplying the response function in time-domain with a Gaussian decay with a width of 45~$\text{cm}^{-1}$ for the two-mode model Hamiltonians and 100~$\text{cm}^{-1}$ for Naphthalene to match the experimental spectrum. To increase the resolution, we padded the damped response functions with zeros such that they reach a length of 2000 grid points. We chose the temperature for the linear spectra to be $T=1~\text{K}$ for the two-mode model Hamiltonians and $T=313~\text{K}$ for naphthalene to match that of the experimental reference.
For the calculation of all 2DRR spectra the grids of $t_1, t_3$, and $t_5$ times can be chosen much smaller than the grids of the $t_2$ and $t_4$ times, since we integrate over $t_1$, $t_3$ and $t_5$ only during the pulse durations. Thus, we chose a grid containing 10$\times$600$\times$10$\times$600$\times$10 (10$\times$100$\times$10$\times$100$\times$10) points reaching up to 100~fs in $t_1$, 3000~fs in $t_2$, 100~fs in $t_3$, 3000~fs in $t_4$ and 300~fs in $t_5$ for the two-mode model Hamiltonians (naphthalene). All spectra were computed with $\omega_L=\omega_{eg}$, i.e., the carrier frequency of the pulses is set to the adiabatic excitation frequency of the system. To broaden the spectra, we multiplied the response function during $t_2$ and $t_4$ by a Gaussian decay with a width of $6~\text{cm}^{-1}$. To increase the resolution, we padded the damped response functions in $t_2$ and $t_4$ with zeros such that they reach a length of 2400 grid points. We chose the temperature for the 2DRR spectra to be $T=1~\text{K}$ for the two-mode model Hamiltonians and $T=293~\text{K}$ for naphthalene. To remove the zero frequency components from the 2DRR spectra, we subtracted the average along $t_2$, $t_4$, and of the whole response function. When calculating the 2DRR spectra of naphthalene, we only considered the first 40 vibrations, which are all vibrations lying below 3000~cm$^{-1}$. 

All two-mode model calculations in this work were performed on an Apple M2 CPU and the calculation of a 2DRR spectrum of a two-mode model Hamiltonian took $\approx 2~\text{min}$ parallelized on 8 cores. The calculations of the 2DRR spectra of naphthalene were performed on a CPU Cluster, where each calculation was parallelized on 64 cores and took $\approx2~$days.

\section{Results and Discussion}

\begin{figure}[pth]
  \includegraphics[]{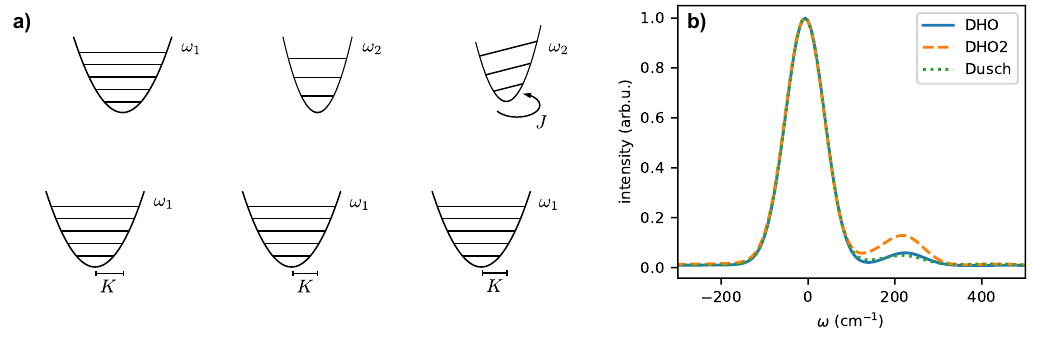}
  \caption{\label{figure2}a) Different models frequently used for the calculation of vibrationally resolved spectra. The DHO model includes only displacements of different vibrational modes between ground and excited state minima (left panel). The second model also includes changes in the frequencies of vibrational modes between ground and excited state (middle panel). The full model further includes Duschinsky coupling between ground and excited state vibrations (right panel). b) Vibrationally resolved linear absorption spectra of three different two-mode model Hamiltonians, two DHO models and one model including Duschinsky coupling.}
\end{figure}

To illustrate the capabilities of our method and to investigate the signature of Duschinsky coupling in 2DRR, we calculated 2DRR spectra of a two-mode model system. Furthermore, we calculated 2DRR spectra of naphthalene based on ab initio data obtained with DFT to verify that the features of Duschinsky coupling observed in the model systems transfer to actual molecules.

\subsection{Two-mode model}

Generally, there are three models that are commonly used when calculating vibrationally resolved spectra including transitions between electronically excited states (Fig.~\ref{figure2}a). First, there is the displaced harmonic oscillator model (DHO), where it is assumed that vibrational modes in the excited state can be displaced, but that the vibrational modes in the ground and excited state are the same, i.e., all vibrations have the same frequency and there is no Duschinsky coupling (Fig.~\ref{figure2}a, left panel). The second model then allows for displacements and changes of the frequencies between ground and excited state vibrations, however, without Duschinsky coupling (Fig.~\ref{figure2}a, middle panel). Finally, the third model includes displacements, frequency changes, and Duschinsky coupling (Fig.~\ref{figure2}a, right panel). In this case, the Duschinsky rotation matrix is defined by
\begin{equation}
    J = \begin{pmatrix}
        \cos\theta & -\sin\theta \\
        \sin\theta & \cos\theta
    \end{pmatrix}
\end{equation}
with the rotation angle $\theta$. The parameters of the models are given in Table~\ref{table2}.

\begin{table}[]
\caption{Parameters for the two-mode model systems.}
\begin{tabular}{Sc|ScScScScSc}\label{table2}
\#model & $\omega_g$ ($\text{cm}^{-1}$) & $\omega_e$ ($\text{cm}^{-1}$) & $K$ & $\theta$ \\ \hline
1       &    $\begin{pmatrix}  150 \\230  \end{pmatrix}$      &     $\begin{pmatrix}  150 \\230  \end{pmatrix}$    & $\begin{pmatrix}  0 \\ 10 \end{pmatrix}$  &  0     \\
2       &    $\begin{pmatrix}  150 \\230  \end{pmatrix}$      &     $\begin{pmatrix}  150 \\230  \end{pmatrix}$    & $\begin{pmatrix}  10 \\ 15 \end{pmatrix}$ &     0  \\
3       &$\begin{pmatrix}  150 \\230  \end{pmatrix}$      &     $\begin{pmatrix}  150 \\230  \end{pmatrix}$    & $\begin{pmatrix}  0 \\ 10 \end{pmatrix}$&     $\pi/6$ 
\end{tabular}
\end{table}

Starting with the two-mode model systems, we compare three different cases. The first two cases are DHO models, where either only one or both modes are displaced in the excited state. The third case then includes displacement of only one mode with additional Duschinsky coupling.

\begin{figure}[pth]
  \includegraphics[]{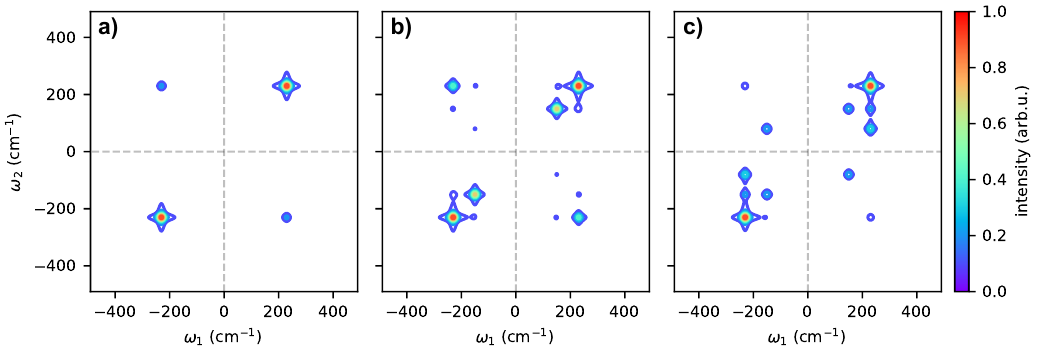}
  \caption{\label{figure3}2DRR spectra of three different two-mode model Hamiltonians, two DHO models and one model including Duschinsky coupling.}
\end{figure}

The linear absorption spectra of these three models are shown in Fig.~\ref{figure2}b. All three linear absorption spectra look qualitatively similar with a large 0-0 peak and a small vibrational progression. Since we have chosen a broadening that is typical for vibrationally resolved linear absorption spectra, the influence of the Duschinsky coupling is not clearly visible. In turn, the width of peaks in 2DRR spectra is usually much smaller, since coherences between vibrational states are intrinsically longer lived than electronic coherences. 

Generally, a 2DRR spectrum can be separated into four quadrants. Since the time-domain signal is real, its Fourier transform satisfies $\sigma_{2DRR}(\omega_1,\omega_2) = \sigma_{2DRR}(-\omega_1,-\omega_2)$. Therefore, the spectrum is such that the first and third quadrants are equal, as are the second and fourth quadrants. Thus, the essential information is contained in the parts of the spectrum with $\omega_2>0$. While the R2 and R4 pathways contribute to the first and second quadrant of the spectrum, the R1 and R3 pathways only contribute to the first quadrant, since they only contain interactions either from the right or from the left during the first four interactions such that there can be no coherences during $T_1$ and $T_2$ with opposite sign. Thus, the first quadrant usually contains the strongest features in 2DRR spectra. The 2DRR spectra of the three model systems are shown in Fig.~\ref{figure3}. Each spectrum is normalized to its own maximum. Consequently, the following comparison concerns the relative distribution of intensity within each spectrum rather than the absolute signal strengths between the three models. The 2DRR spectrum of the first model with a single displaced mode (mode 2) contains only two relevant peaks, one diagonal peak at $(\omega_{g,2}, \omega_{g,2})$ in the first quadrant and a peak at $(-\omega_{g,2}, \omega_{g,2})$ (Fig.~\ref{figure3}a). 

After introducing displacement in the other mode (mode 1), an additional diagonal peak appears at $(\omega_{g,1}, \omega_{g,1})$ (Fig.~\ref{figure3}b). Although we have not introduced Duschinsky coupling between the modes, cross peaks emerge in the first quadrant at $(\omega_{g,1}, \omega_{g,2})$ and $(\omega_{g,2}, \omega_{g,1})$, since the two interactions with the laser between the delay times enable the change of two quanta. Cross peaks in the first quadrant only emerge within the R1 and R3 pathways (see Fig.~\ref{figureA1}), since they require coherences of the first (second) vibrational mode during $T_1$ ($T_2$) on the same side of the density matrix, which is not possible in R2 and R4. Overall, the spectrum in the first quadrant is approximately symmetric with respect to the diagonal. The second quadrant contains only three additional minor peaks compared to the second quadrant of the DHO spectrum. Although there is no additional diagonal peak at $(-\omega_{g,1}, \omega_{g,1})$, two cross peaks appear at $(-\omega_{g,1}, \omega_{g,2})$ and $(-\omega_{g,2}, \omega_{g,1})$. These cross peaks in the second quadrant emerge within the R2 and R4 pathways (see Fig.~\ref{figureA1}). Apart from the cross peaks, only a small difference peak emerges at $(-\omega_{g,1}, \omega_{g,2} - \omega_{g,1}) = (-150 \,\text{cm}^{-1}, 70\,\text{cm}^{-1} )$, which does not have a counterpart at $(\omega_{g,1}-\omega_{g,2}, \omega_{g,1},)$.

If Duschinsky coupling is included, the spectrum changes significantly (Fig.~\ref{figure3}c). In the second quadrant, the small cross peaks, which appeared in Fig.~\ref{figure3}b, disappear. In contrast, the intensity of the difference peak increases. In the first quadrant, there are again two diagonal peaks at the frequencies of the two vibrational modes along with cross peaks between these two modes. Thus, Duschinsky coupling effectively activates the undisplaced (Franck--Condon inactive) mode. Furthermore, there is a larger difference in the intensity of the cross peaks. Although we have reduced the displacement of the second mode, compared to the cross peaks in Fig.~\ref{figure3}b, the intensity of the cross peak at $(\omega_{g,2}, \omega_{g,1})$ has increased while that of the cross peak at $(\omega_{g,1}, \omega_{g,2})$ has decreased in Fig.~\ref{figure3}c. Therefore, the enhanced imbalance between transpose-related cross peaks is a sensitive signature of the intensity redistribution caused by Duschinsky coupling.

In addition, a strong difference peak appears at $(\omega_{g,2}, \omega_{g,2} - \omega_{g,1})$. Generally, difference peaks can only emerge due to the R2 and R4 pathways (Fig.~\ref{figureA2}), since to create such a peak, coherences on opposite sides of the density matrix have to exist during $T_1$ and $T_2$. Also, there cannot exist a difference peak at $(\omega_{g,2} - \omega_{g,1}, \omega_{g,2})$, since the system cannot be in excited vibrational states on both sides of the density matrix during $T_1$. Therefore, the appearance of difference peaks leads to a less symmetric looking 2DRR spectrum. Overall, we observe that Duschinsky coupling leads to more asymmetric spectra.

\subsection{Naphthalene}

\begin{figure}[pth]
  \includegraphics[]{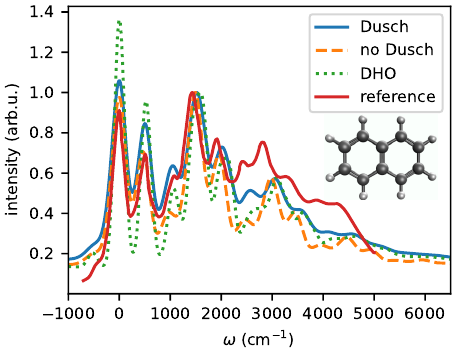}
  \caption{\label{figure4} Vibrationally resolved linear absorption spectra of naphthalene calculated at the (TD)-DFT level of theory with the PBE0 functional. Spectra within the three different models depicted in Fig.~\ref{figure2}a are shown together with an experimental reference taken from \cite{doi:10.1139/v57-152,essd-5-365-2013}.}
\end{figure}

The vibrationally resolved linear absorption spectrum of naphthalene was previously calculated by Benkyi \textit{et al.}.\cite{10.1039/c9cp04178h} They compared the results for different levels of electronic structure theory. As a benchmark they used CC2, where S$_2$ is the first electronically excited state with a high transition dipole moment, and compared it to (TD)-DFT with the PBE0 functional. For TD-DFT, the S$_1$ is the corresponding electronically excited state with a large transition dipole moment. Since the difference between the results of the two electronic structure methods is rather small and the computational demand of (TD)-DFT calculations is lower, we calculate linear and 2DRR spectra of the S$_1$ of naphthalene using (TD)-DFT with the PBE0 functional (for more details see Sec.~\ref{computational}). Vibrationally resolved absorption spectra of the three different models---DHO, uncoupled harmonic oscillators with frequency change, and full Duschinsky rotated harmonic model (Fig.~\ref{figure2}a)---are shown in Fig.~\ref{figure4}, together with an experimental reference taken from Refs.~\citenum{doi:10.1139/v57-152, essd-5-365-2013}. All three spectra were shifted to match the frequency of the 0-0 transition and normalized to the maximum of the experimental spectrum. It is evident that in the spectrum of the DHO model the intensity of the first two peaks is highly overestimated compared to the experiment and compared to models with frequency change and Duschinsky coupling. The intensity of the high frequency tail is underestimated by all three models. 

\begin{figure}[pth]
  \includegraphics[]{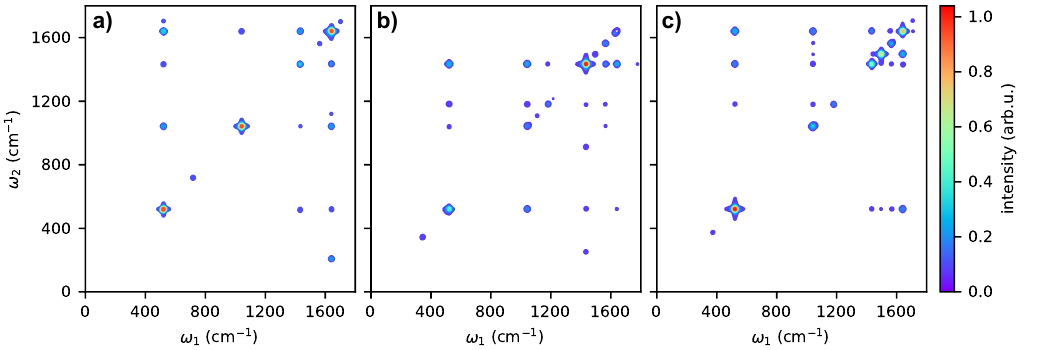}
  \caption{\label{figure5}Normalized 2DRR spectra of naphthalene calculated at the (TD)-DFT level of theory with the PBE0 functional. Each panel is normalized independently to its maximum intensity. a) DHO model, including excited-state displacements while retaining the ground-state frequencies and normal coordinates. b) Curvature-change model, including displacements and different ground- and excited-state frequencies with $J=I$. c) Full harmonic model, including displacements, frequency changes, and Duschinsky rotation.}
\end{figure}

The corresponding 2DRR spectra are shown in Fig.~\ref{figure5}. The DHO spectrum (Fig.~\ref{figure5}a) contains prominent diagonal features at approximately 500, 1000, 1400, and 1600~cm$^{-1}$. These peaks arise when the same vibrational coherence frequency is sampled during both experimental delay periods. A series of off-diagonal cross peaks connects the dominant vibrational frequencies. Their presence does not by itself imply coupling between the normal modes, since even for independent displaced harmonic oscillators, the two Raman coherence periods can involve different modes and can therefore produce cross peaks.

\begin{figure}[pth]
  \includegraphics[]{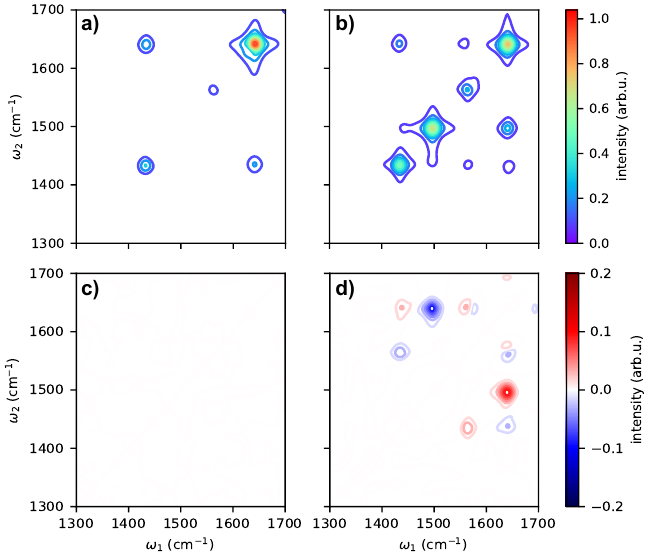}
  \caption{\label{figure6}High-frequency regions of the normalized naphthalene 2DRR spectra and their antisymmetric components. a) DHO model. b) Full model including Duschinsky rotation. c,d) Corresponding antisymmetric components, $\Delta\sigma_\text{2DRR}(\omega_1,\omega_2)=[\sigma_\text{2DRR}(\omega_1,\omega_2)-\sigma_\text{2DRR}(\omega_2,\omega_1)]/2$, for the DHO and full models, respectively.}
\end{figure}

The DHO spectrum is not exactly symmetric with respect to the exchange of the two frequency axes. For example, the cross peak connecting the modes near 500 and 1000~cm$^{-1}$ is more prominent on one side of the diagonal than at its transposed position. Nevertheless, high-frequency cross peaks occur in pairs of nearly identical intensity (Fig.~\ref{figure6}a). To quantify this property, we define the antisymmetric component of the spectrum as
\begin{equation}\label{eq:antisymmetric_2DRR}
    \Delta\sigma_\text{2DRR}(\omega_1,\omega_2)
    =
    \frac{\sigma_\text{2DRR}(\omega_1,\omega_2)-\sigma_\text{2DRR}(\omega_2,\omega_1)}{2}.
\end{equation}
By construction, $\Delta\sigma_\text{2DRR}(\omega_1,\omega_2)=-\Delta\sigma_\text{2DRR}(\omega_2,\omega_1)$, and its value vanishes on the diagonal. The nearly featureless antisymmetric component of the high-frequency DHO spectrum (Fig.~\ref{figure6}c) confirms that the corresponding cross-peak pairs have very similar intensities.

Fig.~\ref{figure5}b shows the result obtained when the calculated excited-state frequencies are retained while the Duschinsky matrix is set to the identity. In contrast to the DHO model, the ground- and excited-state potentials now have different curvatures. Upon electronic excitation, the vibrational wavepacket is therefore not only displaced but also squeezed. This produces clusters of nearby diagonal features, reduces the relative intensities of the peaks near 500 and 1000~cm$^{-1}$, and increases the number of observable cross peaks. In the high-frequency region, the dominant diagonal intensity shifts from the feature near 1600~cm$^{-1}$ in the DHO spectrum to that near 1400~cm$^{-1}$. This behavior can be traced back to the assumption that the Duschinsky matrix is just the identity: If the vibrational frequencies between ground and excited state are different, they might be permuted. If such a change occurs, the Duschinsky matrix compensates this by taking the form of a permutation matrix with respect to these two vibrational modes. However, if the Duschinsky matrix is assumed to be the identity matrix, the missing permutation leads to sudden changes in the intensity of vibrational modes lying close to each other. Therefore, allowing different frequencies of ground- and excited-state vibrations while neglecting Duschinsky coupling may lead to worse results than just taking a DHO model.

The full model also includes the Duschinsky rotation (Fig.~\ref{figure5}c). Positions and clustering of the diagonal features remain broadly similar to those in Fig.~\ref{figure5}b, because both calculations employ the same ground- and excited-state frequencies. However, their intensities are redistributed further. This intensity redistribution between modes modifies the relative strengths of the diagonal peaks, and enhances or suppresses individual cross peaks. In the normalized full spectrum, this redistribution makes the feature near 500~cm$^{-1}$ the strongest diagonal peak relative to the high-frequency features.

The effect is particularly apparent in the high-frequency close-up in Fig.~\ref{figure6}b. Whereas the DHO model produces approximately balanced cross-peak pairs, the full calculation produces markedly different intensities at transposed positions. For example, cross peaks near $(1640,1490)$ and $(1490,1640)$~cm$^{-1}$ give rise to the strongest cross-peak pair in the antisymmetric component shown in Fig.~\ref{figure6}d. Smaller antisymmetric pairs are visible for other combinations of the modes between approximately 1400 and 1650~cm$^{-1}$. These features demonstrate that Duschinsky mixing changes the relative weights of pathways in which the two vibrational frequencies occur in the opposite order during the two Raman coherence periods. Within the comparison of Figs.~\ref{figure5} and \ref{figure6}, the additional cross-peak asymmetry is a sensitive signature of the intensity redistribution caused by Duschinsky rotation.

\section{Conclusions}

The central outcome of this work is a practical framework for evaluating high-order vibronic response functions for general harmonic potential energy surfaces. Using the analytic expressions of the propagator's position space representation allows us to rewrite the required multi-time correlation functions in terms of a multidimensional Gaussian integral, such that the calculation is reduced to standard matrix operations. Our method treats displacements, changes in vibrational frequencies, Duschinsky rotation, and finite-temperature effects without explicitly summing over vibrational states or propagating nuclear wavepackets. Overall, it provides an efficient framework for calculating vibronic response functions of arbitrary order and can be utilized to interpret lineshape effects in high-order multidimensional spectroscopy.

We applied the method to fifth-order 2D resonance Raman spectra of two-mode model systems and naphthalene. The model calculations show that cross peaks can occur even for independent displaced harmonic oscillators and are therefore not, by themselves, evidence of coupling between vibrational modes. Duschinsky rotation instead produces characteristic changes in the relative intensities of diagonal and cross peaks by mixing the normal coordinates of the ground and excited electronic states. In particular, it increases the intensity imbalance between cross peaks related by exchange of the two frequency axes and enhances difference-frequency features. These effects lead to a more pronounced antisymmetric component of the 2DRR spectrum. For naphthalene, the influence of Duschinsky rotation is difficult to identify unambiguously in the broadened linear absorption spectrum but is considerably more apparent in the 2DRR spectrum. The additional spectral dimension and narrower vibrational features make 2DRR spectroscopy particularly sensitive to changes in the normal modes between electronic states.

\section*{Data Availability Statement}

The data underlying this study are openly available on Zenodo at 10.5281/zenodo.21774357. The code that implements the derived closed-form expressions for multi-time correlation functions is available on GitHub at https://github.com/begusic-group/multitime-vibronic.

\begin{acknowledgement}

The authors thank Stefan M\"uller and Tobias Brixner for helpful discussions of high-order spectroscopy. L.N.P. acknowledges a fellowship by the \textit{Fonds der Chemischen Industrie} (FCI).

\end{acknowledgement}

\appendix

\section{Spectra of individual Liouville space pathways}

\begin{figure}[pth]
  \includegraphics[]{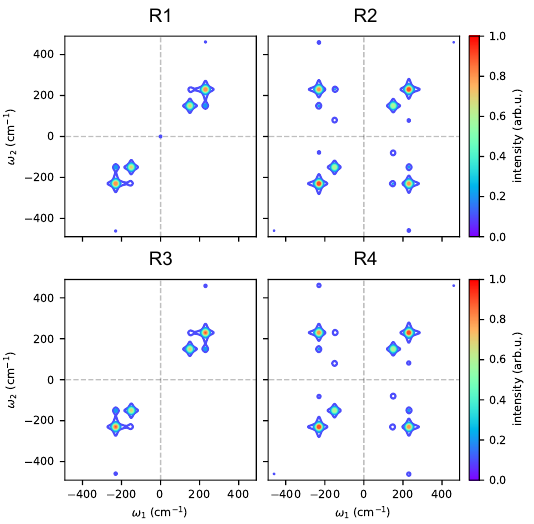}
  \caption{\label{figureA1}2DRR spectrum contributions of the individual Liouville space pathways of the two-mode model Hamiltonian with two uncoupled displaced vibrational modes.}
\end{figure}

\begin{figure}[pth]
  \includegraphics[]{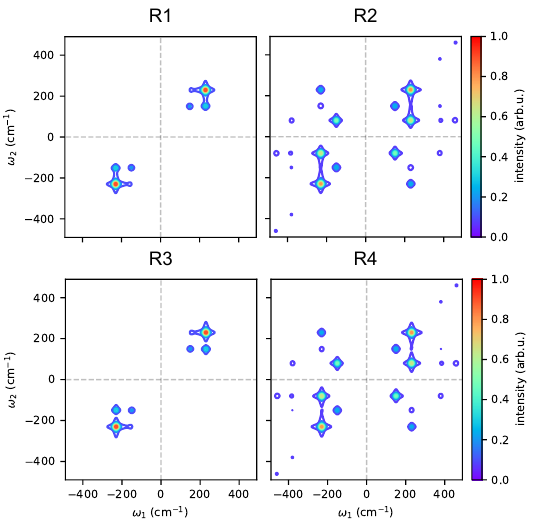}
  \caption{\label{figureA2}2DRR spectrum contributions of the individual Liouville space pathways of the two-mode model Hamiltonian with one displaced vibrational mode, which is coupled by Duschinsky coupling to the second, undisplaced mode.}
\end{figure}


\newpage
\bibliography{bibliography}

\end{document}